\documentclass{IOS-Book-Article}

\usepackage{color,soul}
\usepackage{subcaption}
\usepackage{fancyhdr}
\usepackage{lastpage}
\usepackage{textcomp}
\usepackage{booktabs}
\usepackage{lscape}
\usepackage{tabularx}
\usepackage{makecell}
\usepackage{rotating}
\usepackage{multirow}
\usepackage{listings}
\usepackage[table,xcdraw]{xcolor}
\usepackage{amsmath}
\usepackage{graphicx}
\usepackage{enumerate}
\usepackage{hyperref}
\usepackage{rotating}
\graphicspath{ {images/} }
\usepackage{array, makecell}

\usepackage{mathptmx}
\usepackage{soul}\setuldepth{article}
\def\hb{\hbox to 11.5 cm{}}

\begin{document}

\pagestyle{headings}
\def\thepage{}
\begin{frontmatter}              


\title{Factors Impacting the Quality of User Answers on Smartphones}

\markboth{}{\hb}

\author{\fnms{Ivano} \snm{Bison}}
and
\author{\fnms{Haonan} \snm{Zhao}\thanks{Corresponding Author: Haonan Zhao, PhD Candidate,
email: \url{haonan.zhao@unitn.it}}}

\runningauthor{Ivano Bison and Haonan Zhao}
\address{DISI, University of Trento, Italy.}

\begin{abstract}

So far, most research investigating the predictability of human behavior, such as mobility and social interactions, has focused mainly on the exploitation of sensor data. However, sensor data can be difficult to capture the subjective motivations  behind the individuals' behavior. Understanding personal context (e.g., where one is and what they are doing) can greatly increase predictability. The main limitation is that human input is often missing or inaccurate. The goal of this paper is to identify factors that influence the quality of responses when users are asked about their current context. We find that two key factors influence the quality of responses: user reaction time and completion time. These factors correlate with various exogenous causes (e.g., situational context, time of day) and endogenous causes (e.g., procrastination attitude, mood). In turn, we study how these two factors impact the quality of responses.

\end{abstract}

\begin{keyword}
Human-Machine interaction\sep Context\sep Quality of user answers\sep Diversity awareness
\end{keyword}
\end{frontmatter}
\markboth{May 2023\hb}{May 2023\hb}

\section{Introduction}

Smartphones facilitate pervasive and continuous interaction between humans and machines by enabling timely notifications and data collection, while simultaneously gathering sensor data such as Bluetooth, GPS, and accelerometer. This creates opportunities for smartphone applications to learn how to ask the right questions at the right moment and in the right context, while also verifying the accuracy and consistency of user input. Previous research, as demonstrated in \cite{KD-2020-Bontempelli1, KD-2022-Bontempelli-lifelong}, has explored these capabilities and their potential to establish a symbiotic relationship between humans and machines, wherein machines can learn about every aspect of daily life. This work holds high-impact applications in Artificial Intelligence, Psychology (e.g., the Experience Sampling Method), and the Social Sciences. Additional research on this topic can be found in \cite{H-2014-Wang, giunchiglia2023context, li2022representing}.

However, an outstanding issue remains with user-provided answers as they often fail to meet the necessary quality standards. There are many possible reasons for this, as described in Furnham's work \cite{furnham1986response}, including recall bias, which occurs when participants cannot recall previous events \cite{porta2014dictionary}. These issues can lead to various problems, as highlighted by Schneier \cite{H-2015-Schneier}. In ESM, fixed time intervals are typically used for asking questions, which may result in sending questions at inopportune moments, disrupting people's daily lives and leading to low-quality data collection and, consequently, a limited number of responses. Much research has focused on understanding this phenomenon and improving the response rate by minimizing missed answers, as shown by the works in \cite{H-2020-Van, H-2017-Mishra}, but they did not find which factors can impact the quality of answers.

The goal of this paper is to provide an in-depth study of the parameters that influence the quality of answers. By quality, we refer to a a low number of mistakes. Different individuals possess diverse understandings of questions and context, which leads to variations in the quality of their answers. In this study, we discovered that the correctness of an answer is directly negatively affected by the reaction time, which, in turn, depends on the characteristics of the respondent and the situational context. Thus, controlling the reaction time becomes crucial for establishing meaningful hybrid machine-artificial intelligence.

\section{Methodology}

The reasons for respondents making mistakes when providing answers can be categorized into four main causes: (i) the context in which the answer is provided (e.g., being alone, at university, using social media, on Monday) \cite{S-2010-Lavrakas, S-2021-Wenz}; (ii) the cognitive task involved in the response process \cite{S-1991-Krosnick, S-2013-Lynn}; (iii) a range of psycho-physical attitudes, such as mood, procrastination, response behaviors, and habits \cite{S-2013-Lynn, S-2019-Read}; and (iv) technical problems related to the functioning of the smartphone. These causes are independent of each other but interact over time and across respondents. Based on this understanding, we propose a hypothesis that the quality of answers can be influenced by the reaction time (time elapsed between receiving a question and starting to answer) and completion time (time taken to complete an answer). Both the reaction time and completion time can be affected by the four categories of mistakes described above.

Initially, we hypothesized that longer reaction times could negatively affect answer quality due to memory recall issues. The longer the time elapsed since the request, the greater the risk of memory-related problems and, consequently, making mistakes in the answer. Additionally, we hypothesized that longer completion times could also have a negative influence on answer quality. Furthermore, the combined effect of reaction and completion times could further amplify their negative impact on answer quality.

According to our hypothesis, the primary challenge of our study is to determine how variations in reaction time and completion time contribute to errors in answers. However, solving this problem is complex due to the causal relationship that exists between reaction and completion time in a temporal sequence. To address this issue, we have chosen a causal model \cite{S-1986-Holland, hox2013multilevel}, which establishes a hierarchy that considers the temporal order and logical derivation of variables along the pathway connecting explanatory features and objectives. In this paper, we assess answer correctness by examining the temporal consistency between the respondent's indicated location at home and the GPS position of the phone at the time of notification.

\section{Results and Descriptions}

To achieve the goal of this paper, we analyzed the SmartUnitn 2 database \cite{KD-2021-bison-SU2}, which designed an experiment that lasted four weeks and involved students from the University of Trento. Throughout the experiment, the iLog application  \cite{KD-2014-PERCOM} was used to ask a set of questions diachronically, enabling the selective asking of multiple questions for situational context, which defined in \cite{KD-1993-giunchiglia, KD-2017-PERCOM}. And then collecting sensor data from mobile phones.

\begin{table}[!htpb]
\caption{Chain of errors: Multilevel structural equation model and a Structural Equation model.}
\centering
\begin{tabular}{lll}
\toprule
\multirow{2}{*}{Features}                 & ML-SEM Model\ \ \    & SEM Model       \\ 
                                               & ($\beta$ parameters)                                                       & ($\beta $ parameters)                   \\ \cmidrule(r){1-3}
\textit{M1: Home phone distance (meter)}          &                                  &               \\
Completion time               & 0.2401***                                        & 0.0863***        \\
Reaction time      & 0.4725**                               & 0.0238**     \\
GPS accuracy                              & 0.3439**                                     & 0.0237**     \\ 
Constant                                       & 7.3587                                 & 0.0457     \\      

\textit{M2: Unanswered Questions (count)}         &                                  &            \\
Reaction time           & 0.0252***                                    & 0.8809***   \\ 
Constant                                       & -0.1165*                    & -0.0691  \\    

\textit{M3: Reaction time (minutes)}       &                                  &                      \\
User Characteristics: Procrastination            & 1.0325***                                      & 0.1216***       \\
User Characteristics: Mood                           & 1.4515***                                        & 0.0168*        \\

Event context: Activity                     & 0.5991***                                  & 0.0530***       \\
Study time                   & 1.6248***                                 & 0.1127***          \\
Study time2                  &-0.3105***                          & -0.0716***        \\
Social context: Alone vs not alone                  &-13.5724***                          & -0.1170***       \\
Constant                                       & 18.7450***                      & 0.3596***   \\          

\textit{M4: Completion time (second)}           &                                  &                \\
Pending notification (count)              & -0.4689***                 & -0.0946***          \\
Question delivery delay time (sec.) & -0.0012**                                   & -0.0212*          \\
Event context: Activity                     & -0.2697***                                   & -0.1784***       \\
Social context: Alone vs not alone                      & -2.521***                   & -2.521***          \\
Study time                   & -0.1818***             & -0.0844***         \\
Study time2               & 0.0203***                & 0.0357***  \\ 
User Characteristics: Procrastination            & 0.0766***                                     & 0.0644***      \\
Constant                                       & 13.7792***                     & 1.7036***        \\
 \bottomrule
\end{tabular}
\label{H3}
\end{table}

\vspace{-0.45cm}
\noindent
\textit{\scriptsize Note: ``P-value”, providing a measure of the statistical significance of that feature compared to the reference feature. Here ``*” means p\textless 0.1, namely, relatively weak (statistical) significance; ``**” is  p\textless 0.05, a strong significance; ``***” is  p\textless 0.01, a very strong significance.}

\vspace{0.4cm}
\normalsize
We study the correctness of the user answers by building a causal model able to capture both the direct and indirect effects of the various features. We compute the correctness of answers by comparing the distance from home, as computed by the device via the GPS information, any time the respondent declares she is at home. Notice that each participant was asked to provide their home address, we then used google maps were compute their home GPS coordinates; Google maps were used to calculate the GPS coordinates of the house. 

The results are reported in Table \ref{H3}. Horizontally we have a path model compose to four models (\textit{M1-M4}). The variable written in italic near the name of the model is the dependent variable, the variables listed below are the input variables. In both models, we report the correlation (i.e., ``Coeff $\beta $") between input and dependent variable, for each model in \textit{M1-M4}. Out of the two models, the first is a Multilevel Structural Equation Model (ML-SEM) \cite{S-2004-Rabe} , and a Structural Equation Model (SEM) \cite{S-2007-Rabe}. Both models (MSEM, SEM) support the idea of a chain of event on answer quality. In fact, the input variables are (mostly strongly) relevant and the fit indices show that the model fits the observed data very well (RMSEA= 0.022; SRMR=0.013; chi2=95.41 (20); $R^2= 0.127$). Moreover, the model explains 13.0\% of fluctuation in answer quality. 

In \textit{M1}, completion and reaction time correlate positively with ``Home $\Leftrightarrow$ phone distance" controlling by the smartphone accuracy. Their interaction has a series of negative consequence on answer quality. In fact, a longer reaction time has two effects. The first a direct effect on increase the distance between ``Home $\Leftrightarrow$ phone”. The second indirect, thru the increase of the ``pending notifications" \textit{(M2)}, which, in turn, induce a shorter completion time which in turn decreases the error. This is not a contradictory result, but the twofold way in which an error can appear.

Our results proof that subjective annotations present a certain degree of error due to both exogenous and endogenous factors \textit{(M3, M4)} affecting the quality of responses. Context history, cognitive ability, attention, effort, motivation, burden, procrastination, mood, and technical problems can play a role in terms of raising the probability of stopping the interaction with the machine, of not compliance with the interaction protocol, of a decrease in the level of attention and, consequently, they can cause a decrease of the accuracy of answers.

\section{Conclusion}

This paper has investigated the effects of various factors on the quality of answers, in terms of missed answers as well as their correctness. The key results of this paper are of two kinds. The first kind is that in the future the researcher's attention should be placed on several factors related to: (a) controlling the situational and temporal context to find the best moment for administering a notification; (b) focusing on the human-machine interaction not only on the layout of the apps, but on the structure and order of the response alternatives, the ease of filling in, and finally on the support of the machine to help respond so as to reduce the response time and improve its quality. The second kind of results is related to the cognitive and more generally psychosocial traits of the respondents. It is clear that not all subjects are cooperative and follow the research protocols carefully. In the future, it will be a matter of finding what and how cognitive factors act differently and how to extrapolate their data and replace missing data from the few and fragmented data provided. The future work will focus on how to minimize the reaction time by establishing the best moment for asking a question.

\section*{Acknowledgments}
The research by Ivano is funded by the European Union's Horizon 2020 FET Proactive project “WeNet – The Internet of us”, grant agreement No 823783. The work by Haonan receives funding from the China Scholarships Council (No.202107820038).

\newpage

\bibliographystyle{splncs04}
\bibliography{bibliography,HumanInLoop,KnowDive,Sociology}

\end{document}